\documentclass[fleqn,11pt]{article}

\usepackage{amsfonts,amsmath,amssymb}
\usepackage{latexsym}

\usepackage{amsfonts,amssymb,cite}
\usepackage{graphicx}

\def\noi{\noindent}

\newcommand{\Title}[1]{\noi {{\Large\bf #1}}\\[1ex]}

\def\Aunames#1{\noi{\bf #1}}
\def\au#1{${}^{#1}$}
\def\Addresses#1{\medskip\noi \protect
	\begin{description}\itemsep -3pt {\it #1} \end{description}}
\def\adr#1#2{\item[${}^{#1}$]{\it #2}}

\newcommand{\Abstract}[1]{\vskip 2mm \begin{center}
        \parbox{16.4cm}{\small\noi #1} \end{center}\medskip}

\def\email#1#2{\footnotetext[#1]{e-mail: #2}\addtocounter{footnote}{1}}

\def\nq{\hspace*{-1em}}
\def\nqq{\hspace*{-2em}}
\def\nhq{\hspace*{-0.5em}}

\def\qq{\qquad}
\def\cm{\hspace*{1cm}}

\def\wide{\mbox{$\dst\vphantom{\int}$}}

\def\ten#1{\mbox{$\times 10^{#1}$}}
\usepackage{color}

\def\Acknow#1{\subsection*{Acknowledgments} #1}

\def\Jl#1#2{#1 {\bf #2},\ }

\def\ApJ#1 {\Jl{Astroph. J.}{#1}}
\def\CQG#1 {\Jl{Class. Quantum Grav.}{#1}}
\def\DAN#1 {\Jl{Dokl. AN SSSR}{#1}}
\def\GC#1 {\Jl{Grav. Cosmol.}{#1}}
\def\GRG#1 {\Jl{Gen. Rel. Grav.}{#1}}
\def\IJMPD#1 {\Jl{Int. J. Mod. Phys. D}{#1}}
\def\JETF#1 {\Jl{Zh. Eksp. Teor. Fiz.}{#1}}
\def\JETP#1 {\Jl{Sov. Phys. JETP}{#1}}
\def\JHEP#1 {\Jl{JHEP}{#1}}
\def\JMP#1 {\Jl{J. Math. Phys.}{#1}}
\def\NPB#1 {\Jl{Nucl. Phys. B}{#1}}
\def\NP#1 {\Jl{Nucl. Phys.}{#1}}
\def\PLA#1 {\Jl{Phys. Lett. A}{#1}}
\def\PLB#1 {\Jl{Phys. Lett. B}{#1}}
\def\PRD#1 {\Jl{Phys. Rev. D}{#1}}
\def\PRL#1 {\Jl{Phys. Rev. Lett.}{#1}}

\def\al{&\nhq}
\def\lal{&&\nqq {}}
\def\eq{Eq.\,}
\def\eqs{Eqs.\,}
\def\beq{\begin{equation}}
\def\eeq{\end{equation}}
\def\bear{\begin{eqnarray}}
\def\bearr{\begin{eqnarray} \lal}
\def\ear{\end{eqnarray}}
\def\earn{\nonumber \end{eqnarray}}
\def\nn{\nonumber\\ {}}

\def\nnn{\nonumber\\ \lal }

\def\yy{\\[5pt] {}}
\def\yyy{\\[5pt] \lal }
\def\eql{\al =\al}
\def\eqv{\al \equiv \al}

\def\dst{\displaystyle}

\def\fracd#1#2{{\dst\frac{#1}{#2}}}

\def\Half{{\fracd{1}{2}}}

\def\then{\ \Rightarrow\ }

\def\eqn#1{\eq\eqref{#1}}
\def\rf{\eqref}

\def\sph{spherically symmetric}
\def\ssph{static, spherically symmetric}

\begin{document}
\twocolumn[


\Title{Gravity assist as a test of relativistic gravity}

\Aunames{S. V. Bolokhov,\au{a;1} K. A. Bronnikov,\au{a,b,c;2} and M. V. Skvortsova\au{a;3}}

\Addresses{\small
\adr a {Institute of Gravitation and Cosmology, Peoples' Friendship University of Russia (RUDN University),\\ 
		ul. Miklukho-Maklaya 6, Moscow 117198, Russia}
\adr b {Center for Gravitation and Fundamental Metrology, VNIIMS, 
		Ozyornaya ul. 46, Moscow 119361, Russia}
\adr c {National Research Nuclear University ``MEPhI'', 
		Kashirskoe sh. 31, Moscow 115409, Russia}
        }
        

\Abstract{We consider the gravity assist maneuver, that is, a correction of spacecraft motion at its 
		passing near a planet, as a tool for evaluating the Eddington post-Newtonian parameters 
		$\beta$ and $\gamma$, characterizing vacuum spherically symmetric gravitation fields in 
		metric theories of gravity. We estimate the effect of variation in $\beta$ and $\gamma$ 
		on a particular trajectory of a probe launched from the Earth's orbit and passing closely near
		Venus, where relativistic corrections slightly change the impact parameter of probe scattering 
		in Venus's gravitational field. It is shown, in particular, that a change of $10^{-4}$ in 
		$\beta$ or $\gamma$ leads to a shift of about 50 km in the probe's aphelion position.}

] 
\email 1 {boloh@rambler.ru}
\email 2 {kb20@yandex.ru} 
\email 3 {milenas577@mail.ru}

\section{Introduction}

  By a gravity assist (GA) maneuver, we mean a way to adjust a spacecraft trajectory using the gravitational 
  field of a massive celestial body (a planet) when the spacecraft enters the sphere of gravitational influence 
  of this planet (the Hill sphere). The goal of this maneuver is to increase or decrease the spacecraft 
  velocity in the Sun's (quasi-inertial) reference frame (RF), saving fuel by borrowing kinetic energy from
  the planet's orbital motion. In astronautics, GA has been successfully used for more than 50 years in
  various space missions in the Solar System. A good introduction to the GA problems including many 
  technical and computational aspects can be found in \cite{ceri}, see also~\cite{Kaplan, bartlett}.
  
  Besides fuel economy and technical convenience for tasks of a particular mission, one more important 
  aspect of GA is a high sensitivity of the spacecraft trajectory to variations of its initial parameters 
  of motion such as the impact parameter~\cite{yef-20}. Due to this, it was emphasized that GA can serve 
  as one of possible tools for revealing the difference between Newtonian gravity and Einstein's general relativity 
  (GR) \cite{yef-21}.
  
  In this paper, we develop this idea and demonstrate that GA can be regarded as a tool for quite a precise test 
  of metric theories of gravity, whose difference from GR can be formulated in a suitable approximation in 
  terms of the parametrized post-Newtonian (PPN) formalism \cite{nord-69, will-93, will-14}. 
  In the case of a spherically symmetric gravitational field of a central body, relevant are two well-known 
  PPN parameters $\beta$ and $\gamma$, called the Eddington parameters \cite{edd-22}, that slightly modify 
  the Schwarzschild metric \cite{weinberg}. In GR we have $\beta=\gamma=1$. The influence of these 
  parameters is greatly amplified by GA that governs the probe's subsequent motion, in particular, the
  values of $\beta$ and $\gamma$ affect such measurable quantities as the coordinates of probe's destination 
  points and travel times. Therefore, GA can in principle contribute to the current estimates and restrictions 
  on $\beta$ and $\gamma$ from the astronomical observations in the Solar System. At present, deviations 
  of these parameters from unity are estimated as $\sim 10^{-4}$ \cite{will-14}.
  
  Remarkably, unlike astronomical events, a gravitational maneuver is an	initiated and technically controlled 
  event with a desired accuracy and	calibration, which in principle opens the way for testing theories of 
  gravity in an active manner rather than on the base of passive astronomical observations.  
    
  Needless to say that real calculations of spacecraft motion take into account the elliptic nature of planetary
  orbits and perturbations due to various factors acting in space. Therefore, such calculations are impossible 
  in the sense of exact solutions to the equations of motion and require either direct numerical integration 
  or approximate analytical methods. In our paper we follow the latter way and describe the geodesic motion 
  up to some negligible terms using a realistic approximation from the viewpoint of the currently achievable
  measurement accuracy. Also, for our demonstration purposes it is sufficient to adhere to \sph\ metrics
  and to assume circular planetary orbits. 
    
  As a particular problem to be considered, we take the probe flight model described in \cite{yef-20}. The 
  probe starts from the Earth's orbit and approaches Venus, its flyby near Venus is implemented as a gravity 
  assist maneuver, and, as a result, the probe flies far beyond the Earth's orbit. Eventually, the influence 
  of relativistic corrections and the GA lead to a measurable shift of an arrival point, which is here taken 
  to be the aphelion of the probe's orbit.   

  A note on the notations: we mostly use small letters for parameters of motion in the Sun's RF, capital
  letters for the same in Venus's RF, and the indices 1 and 2 for probe motion before and after the assist, 
  respectively. 
  
  In Section 2 we consider the general formalism of motion in \sph\ space-times and verify the correctness
  of the resulting formulas by reproducing the known expression for the post-Newtonian perihelion shift
  with arbitrary $\beta$ and $\gamma$. Section 3 is devoted to a particular kind of probe orbits with
  GA at Venus, and Section 4 is a brief conclusion.  

\section{Orbital motion and perihelion shift}
\subsection{Post-Newtonian approximation}

  Consider the general \ssph\ metric and its post-Newtonian (PN) representation,
\bear               \label{sph}
	ds^2 \eql f(r) dt^2 - g(r) dr^2 - r^2 d\Omega^2
\yy               \label{sph-PN}
	    \eqv  \Big(1 - \frac{2\mu}{r} + \frac{2(\beta - \gamma) \mu^2}{r^2}\Big) dt^2 
\nnn \cm\cm	    
			       - \Big(1 + 2\gamma \frac{\mu}{r} \Big) dr^2 - r^2 d\Omega^2,
\ear
  where $d\Omega^2 = d\theta^2 + \sin^2\theta d\varphi^2$,  
  the constants $\beta$ and $\gamma$ are PN coefficients, both equal to 1 in the case of GR, 
  and different for different metric theories of gravity; $\mu = GM/c^2$, $M$ being the mass of the 
  gravitating center (so that $r_g=2\mu$ is its Schwarzschild gravitational radius).
  
  We will deal with geodesic motion in the equatorial plane, $\theta = \pi/2$, so that in the general 
  metric \rf{sph} we have two integrals of the geodesic equations and the 4-velocity normalization 
  condition $u_\mu u^\mu =1$ (the dot denotes $d/ds$, and the displacement $ds$ coincides with
  that of proper time of a moving body, $d\tau$):
\bearr              \label{t-dot}
		\dot t = E/f(r),   
\yyy                \label{phi-dot}
		\dot \phi = L/r^2, 
\yyy                \label{norm}
		E^2/f(r) - g(r) {\dot r}^2 - L^2 /r^2 = 1.
\ear		 
  where the constants $E$ and $L$ are conserved energy and angular momentum, respectively, per 
  unit mass of the moving body. If the geodesic motion is finite and takes place between a 
  minimum radius $r_-$ and a maximum radius $r_+$, then these $r_\pm$ can be found from the 
  condition $\dot r =0$, which, being substituted to \rf{norm}, leads to
\beq                \label{r_pm}    
 		\frac{L^2}{r_+^2} = \frac{E^2}{f_+} -1, \quad\
 		\frac{L^2}{r_-^2} = \frac{E^2}{f_-} -1, 
\eeq 
  where $ f_\pm = f(r_\pm)$. This allows for expressing the constants $E$ and $L$ in terms of $r_\pm$:
\beq                  \label{ELpm}
		E^2\! = \frac {f_+ f_- (r_+^2\! - \! r_-^2)}{f_- r_+^2 - f_+ r_-^2},\quad
		L^2\! = \frac {(f_+\! - \! f_-) r_+^2  r_-^2}{f_- r_+^2 - f_+ r_-^2}.
\eeq  

  Comparing $\dot\phi$ in \rf{phi-dot} and $\dot r$ expressed from \rf{norm},
  we can exclude $ds$, which results in 
\beq                \label{dphi1}
           \frac{d\phi}{dr} = \frac{L}{r^2} \frac{\sqrt{f(r)g(r)}}{\sqrt{E^2 - f(r)(1+ L^2/r^2)}}.
\eeq  
  Integration of \rf{dphi1} makes it possible to find the total azimuthal angle $\phi$ covered at 
  test body motion between two specified values of $r$ (on segments where $r$ changes 
  monotonically), provided that $E$ and $L$ are known. 

  Equations \rf{t-dot}--\rf{dphi1} are valid for an arbitrary metric \rf{sph}. However, dealing with 
  the post-New\-to\-nian metric \rf{sph-PN} applied to the Solar system, it makes sense 
  to fix the necessary accuracy of further calculations. Specifically, if $M=M_\odot$ 
  is the solar mass, then the Schwarzschild radius is $r_g = 2\mu \sim 1$ km. The orbital radii 
  are of the order $\sim 10^8$ km, while the Earth's orbital velocity $v \sim 30\,{\rm km/s} 
  \sim 10^{-4}$ in units where $c=1$ shows the magnitudes of all other relevant velocities. Thus,
\bearr                     \label{orders}
                   \frac \mu r \sim 10^{-8}, \qq L \sim 10^4\,{\rm km}, 
\nnn                   
                   \frac{L^2}{r^2} \sim v^2 \sim E - 1 \sim 10^{-8}.
\ear    
  It proves to be reasonable to calculate angles up to $10^{-8}$, and in all expressions we will 
  neglect any terms that make smaller contributions. 
  
  Substituting $f(r)$ and $g(r)$ from \rf{sph-PN} to \rf{dphi1}, we obtain 
\beq                   \label{dphi2} 
       d\phi = \frac {dr}{r^2}\, \frac{L (1 - \mu /r) (1 + \gamma\mu/r)}
                 					{\sqrt{2{\tilde E}+ 2\dfrac \mu r - \dfrac{k+L^2}{r^2} 
                 						- \dfrac{2\mu L^2}{r^3}}},
\eeq   
  where we have denoted
\bearr                  \label{Em}
                 E_m = E - 1 = O(10^{-8}), \quad\  2{\tilde E} = 2E_m + E_m^2,
\nnn                 
                  k = 2\mu^2 (\beta-\gamma).
\ear  
  The term with $1/r^3$ in \rf{dphi2} is eliminated by putting $r = R - \mu$, which results in
\beq                   \label{dphi3}
			d\phi = \frac {L\, dr \, \big[ 1 + (\gamma + 1)\mu/R\big]}
			               {R^2 \sqrt{2\tilde E + 2\dfrac\mu R + \dfrac{2\mu^2 -L^2 - k}{R^2}}},
\eeq  
  where we have neglected some contributions of the order $O(10^{-16})$. This expression is 
  integrated in elementary functions by putting $R = 1/x$: 
  the angle covered between two radius values $r_1$ and $r_2$ is  
\bearr           \label{phi}
		\phi = \bigg(1 + \frac{2\mu^2 -k}{2 L^2}\bigg) 
				\int_{r_1}^{r_2} \frac{- dx}{\sqrt{A_1 + B_1 x -x^2}}
\nnn  \cm
				+ (\gamma + 1) \mu \int_{r_1}^{r_2} \frac{-x\,dx}{\sqrt{A_1 + B_1 x -x^2}}
\nnn \nq\,
	     =\! \Bigg[\bigg(1 + \frac{2\mu^2{-} k}{2 L^2} + \frac {(\gamma{+}1)\mu B_1}{2}\!\bigg)	
		     			\arccos \frac{2 x -B_1}{\sqrt{4A_1\! +\! B_1^2}}
\nnn \cm		     			
		     + (\gamma+1)\mu \sqrt{A_1 + B_1 x -x^2}\Bigg]_{r_1}^{r_2},
\ear  
  where, by our substitutions, $x = 1/(r+\mu)$, and  
\beq            \label{A1B1}
		A_1 = \frac {2E_m + E_m^2}{L^2 + k - 2\mu^2}, \quad 
		B_1 = \frac {2\mu}{L^2 + k - 2\mu^2}.
\eeq  

  Now, with \rf{phi}, we are ready to calculate the anomalous perihelion shift of a planetary orbit 
  due to relativistic gravity. To obtain this shift per revolution, we must calculate the integral \rf{phi}
  between $r_1 = r_-$ (perihelion) and $r_2 = r_+$ (aphelion) and multiply it by two: 
  $\phi = 2\int_{r_-}^{r_+} d\phi$.
    
  We first of all notice that $dr/d\phi =0 \then dx/d\phi =0$ at $r = r_\pm$ (since these are extremal
  points of the function $r(\phi)$), therefore, according to \rf{dphi3}, $A_1 + B_1 x -x^2 =0$ at these 
  points (with necessary accuracy), and thus only the $\arccos$ term in \rf{phi} contributes to $\phi$. 
  Next, after some simple algebra, taking into account the notations \rf{Em}, \rf{A1B1} and the 
  expression \rf{ELpm} for $L^2$, and again neglecting terms of the order $\sim 10^{-16}$, we find that  
\bearr                       \label{eta-pm}
		\frac{2x(r_\pm) -B_1}{\sqrt{4A_1 + B_1^2}} = \mp 1\ 
\nnn		
		\then \
		 \arccos \frac{2 x -B_1}{\sqrt{4A_1 + B_1^2}}\bigg|_{r_1}^{r_2} = \pi,
\ear
  and we conclude that
\beq
		\phi = 2\pi \bigg(1 + \frac{2\mu^2 -k}{2 L^2} + \frac{(\gamma+1)\mu B_1}{2}\bigg).
\eeq  
  It is important that \eqn{eta-pm} is valid with appropriate accuracy for 
  relativistic orbits with arbitrary PN coefficients $\beta$ and $\gamma$.

  Since in Newtonian theory we have $\phi = 2\pi$, the anomalous shift per revolution is
\bearr
          \Delta\phi = \phi - 2\pi = \pi \bigg(\frac{2\mu^2 -k}{L^2} + (\gamma+1)\mu B_1\bigg)
\nnn \qq          
                        \approx \frac{\pi\mu (r_+ +r_-)}{r_+ r_-}(2 + 2\gamma - \beta), 
\ear
  where the last approximate equality is obtained by substitution of $B_1$ from \rf{A1B1} and 
  $L^2$ from \rf{ELpm} and preserving only the main order of magnitude.  
  Recalling that the major semiaxis $a$ of an ellipse and its eccentricity $e$ are related to 
  $r_+$ and $r_-$ by 
\[
              a (1-e^2) = \frac{2 r_+ r_-}{r_+ + r_-},
\]     
  we can finally write
\beq
		\Delta\phi = \frac{2 \pi\mu}{a(1-e^2)}(2 + 2\gamma - \beta),
\eeq  
  in agreement with \cite{weinberg} (see also references therein).
  In particular, in GR, where $\beta = \gamma =1$, we obtain the well-known expression 
  $\Delta\phi = 6 \pi GM/[a(1-e^2)]$, leading to the famous value of $43''$ per century for
  the planet Mercury.

\subsection{Newtonian orbital motion}  
  
  For an arbitrary point of the path we should use the full expression \rf{phi},
  separately for each monotonicity range of $r(\phi)$. To obtain the relativistic correction to
  $\phi$ at the same point, one should subtract the corresponding expression obtained in 
  Newtonian gravity. 
 
  The Newtonian equations of motion of a test body in the gravitational field of an attracting 
  center of mass $M$, moving in the equatorial plane $\theta = \pi/2$ in the spherical coordinates 
  ($r, \theta, \phi$), read
\beq         \label{eq-N}
		\ddot r - r {\dot\phi}^2 = - \frac{GM}{r^2}. \qq r\ddot\phi - 2\dot r \dot\phi =0,  
\eeq  
  where the dot denotes a time derivative, $d/dt$. Equations \rf{eq-N} have the integrals
\beq          \label{int-N}
		r^2 \dot\phi = L_N, \qq    
		\Half \dot r{}^2 + \frac{L_N^2}{2 r^2} - \frac{GM}{r} = E_N,
\eeq  
  where $L_N$ and $E_N$ have the meaning of the Newtonian angular momentum and total 
  energy, respectively, per unit mass of the test body. Excluding $dt$ from the two equations 
  \rf{int-N}, we obtain
\beq
		\bigg(\frac{d\phi}{dr}\bigg)^2 = \frac {L_N^2}{r^4 \big(2E_N + 2GM/r - L_N^2/r^2\big)},
\eeq  
  whence it follows
\bear                 \label{dphi-N}
		\pm \phi \eql \int \frac{dr}{r^2 \sqrt{A_N + B_N/r - 1/r^2}} 
\nn		
			\eql -\int \frac{d\xi}{\sqrt{A_N + B_N \xi  - \xi^2}},
\ear
  where 
\beq            \label{ANBN}
		\xi = \frac 1r, \quad  A_N = \frac{2E_N}{L_N^2}, \quad   B_N = \frac {2GM}{L_N^2}. 
\eeq  
  Integrating \rf{dphi-N} between some radii $r_1$ and $r_2$, we obtain
\beq           \label{phi-N}
		\pm \phi =  \arccos \frac {2/r - B_N}{\sqrt {4 A_N + B_N^2}}\Bigg|^{r_2}_{r_1}.
\eeq  
  It is the expression \rf{phi-N} that should be subtracted from \rf{phi} to obtain the relativistic 
  correction, and a specific example of such a calculation will be considered in the next section. 
  Let us also present some useful formulas for such Kepler elliptic orbits (see, e.g., \cite{LL-meh}):
\bearr               \label{Kep}
		r(\phi) = \frac{p}{1-e\cos \phi}, \quad\
           r_{\pm} = \frac{p}{1 \pm e} = a(1\pm e), 
\nnn   
           p = \frac{L_N^2}{GM}, \quad  
           e = \frac{r_+ - r_-}{r_+ + r_-} = \sqrt{1 + \frac{2 E_N L_N^2}{(GM)^2}},
\ear  
  where $e$ is the orbit eccentricity, $p$ is the focal parameter (which in our case coincides 
  with Venus's orbital radius, equal to OB = OC in Fig.\,1), and $a = (r_+ + r_-)/2$ is the 
  major semiaxis of the ellipse.   
  
  It is of interest that at the intersection points B and C of the two orbits, the horizontal ($x$) 
  component $v_x$ of the probe orbital velocity coincides with that of Venus:
\beq        \label{v_x}
		v_x = v_V = \pm \sqrt{\frac{GM}{p}},
\eeq  
  and this circumstance makes easier our further calculations. This relation may be proved as
  follows. Since $\vec r = (x, y) = (r \cos\phi, r\sin\phi)$, both $v_x = \dot x$ and 
  $v_y = \dot y$ are easily found from the expression of $r(\phi)$ in \rf{Kep}:
\bearr
			v_x = \frac {d [r(\phi) \cos\phi]}{d\phi}\dot \phi = - \frac {L_N}{p} \sin\phi,
\nnn
			v_y = \frac {d [r(\phi) \sin\phi]}{d\phi}\dot \phi = - \frac {L_N}{p} (\cos\phi - e),
\ear 
  where we have used that $\dot\phi= L_N/r^2$. At points B and C where $x =0,\ y = \pm p$,
  and $\phi = \pi/2$ or $3\pi/2$, respectively,
  we have 
\beq           \label{v_xy}  
			v_x = \mp \frac {L_N}{p}, \qq v_y = - \frac {e L_N}{p} = - \dot r(C),
\eeq  
  where the last equality follows from $r\dot r = x\dot x + y\dot y$, while at points B and C we 
  have $x=0,\ y = \pm p= \pm r$. Then \eqn{v_x} follows from the standard expression for the 
  focal parameter of Kepler elliptic motion, $p = L_N^2/(GM) \then L_N = \sqrt{GMp}$. 
  (The latter can be verified by comparing expressions for the conserved quantity $E_N$ 
  in \rf{int-N} applied to $r = r_+ = r_E$ and $r = p$, with substitution of $\dot r$ at $r=p$ and 
  the equality $r_+ = p/(1-e)$.)
  
  Curiously, it happens that the specific angular momentum $L_N$ is the same for the probe 
  and the planet whose circular orbit is crossed by the probe at $\phi = \pi/2$ and $\phi = 3\pi/2$.
   
  Let us note that the expression \rf{phi-N} cannot be obtained directly from \rf{phi} by using the 
  Newtonian approximation of the metric \rf{sph} 
\beq               \label{sph-N}
	    ds^2 =  \Big(1 - \frac{2\mu}{r}\Big) dt^2 - dr^2 - r^2 d\Omega^2,
\eeq
  which corresponds to \rf{sph-PN} with $\beta = \gamma =0$, leading to $k=0$ in \eqs
  \rf{dphi2}--\rf{A1B1}. Indeed, comparing \rf{phi} with $\beta=\gamma=k=0$ with \rf{phi-N},
  we see many differences: an addition to unity in the factor before the arccosine, the nonzero second
  term, and distinctions of $A_1$ and $B_1$ from $A_N$ and $B_N$ that play similar roles in 
  the integration. A reason is, at least partly, in the different meaning of similar derivatives in 
  relativistic and Newtonian equations (a particle's proper time versus Newtonian absolute time), 
  so that, for example, the Newtonian angular momentum $L_n$ is different from $L$.
  More generally, while using \rf{sph-N}, we are still dealing with 4D Riemannian space-time,
  whose geometry requires non-Newtonian laws of motion. The correspondence principle, 
  the requirement that relativistic gravity becomes Newtonian in the case of weak gravity and small
  velocities, works precisely only at zero velocity of a test particle at large $r$ in the metric \rf{sph-N}.
  As we see, at nonzero velocities there emerge corrections comparable to post-Newtonian ones. 
    
\section{Gravity assist with Venus}

\begin{figure}
\centering
\includegraphics[width=8cm]{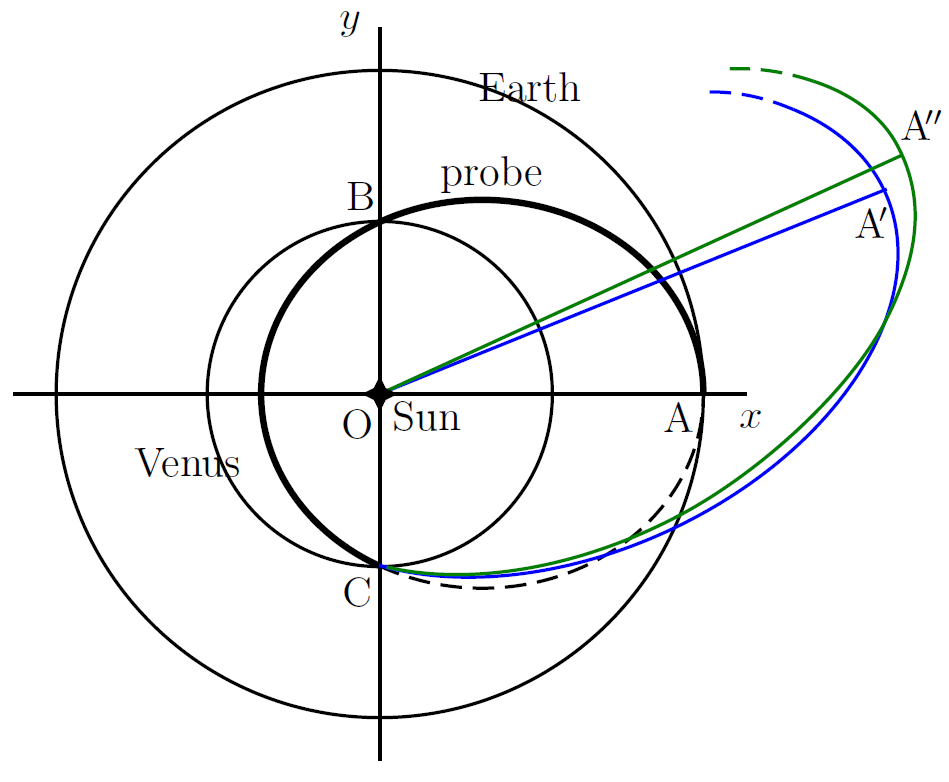} 
\caption{\small 
		The orbits of Earth, Venus and the probe. The arcs $\rm CA'$ and $\rm CA''$ show the 
		probe motion after scattering against Venus with slightly different impact parameters.}
		\label{fig1}
\end{figure}  

   As an example of the effect of different metric theories of gravity on the results of a gravity 
   assist, we will consider such an assist by Venus at point C according to Fig.\,1.
   We assume that Earth and Venus are moving in circular orbits with radii $r_E$ and $r_V$
   and orbital velocities $v_E$ and $v_V$, respectively; the probe under consideration is launched 
   at point A of Earth's orbit with a velocity $v_0 < v_E$ so that its perihelion P is located 
   inside the orbit of Venus, $r_P < r_V$. The initial conditions are
\beq
		r = r_E, \quad  \dot r =0, \quad \phi = 0, \quad \dot\phi = v_0/r_E.
\eeq   
   
  The relevant numerical data to be used in the calculations and some calculated 
  quantities are 
\bearr  		\label{N_data}
		r_E = 1.495878159\ten{8}\,{\rm km},
\nnn		
		r_V = 1.082076791\ten{8}\,{\rm km},
\nnn		
		r_P = 0.847605588\ten{8}\,{\rm km},
\nnn		
		v_0 = 25.336\,{\rm km/s} = 0.0000845118\ ({\rm if}\ c=1),  
\nnn
		v_V = 35.02530368\,{\rm km/s}.		  
\ear

  We will study the probe motion in the framework of Newtonian gravity, while the small relativistic 
  corrections affect only the gravity assist parameters. There influence reduces to the fact that 
  the probe's impact parameter $b$ at its scattering against Venus depends on the correction 
  $\Delta\phi$ of the azimuthal angle $\phi$ at point C, considered in Sun's reference frame (Fig.\,1).
  
\subsection{The relativistic effect}  

  Let us calculate the relativistic correction $\Delta\phi$, the difference between the expressions  
  \rf{phi} and \rf{phi-N} for a fixed value of $r$. For the probe motion from point A to point C
  (or any other point between the perihelion P and a would-be return to A) we can write
\bearr                    \label{Dphi_C}
		\Delta\phi = \frac{\pi\mu}{a(1-e^2)}(2 + 2\gamma - \beta) 
\nnn		
		+\Bigg[\bigg(1 + \frac{2\mu^2 -k}{2 L^2} 
		+ \frac {(\gamma+1)\mu B_1}{2}\bigg) \arccos \eta_1
\nnn 
	     + (\gamma\! + \! 1)\mu \sqrt{A_1\! + B_1 x -\! x^2}\Bigg]_{r_P}^{r}
		    \nq  - \arccos \eta_N \bigg|_{r_P}^r,
\ear
  where $x = 1/(r+\mu)$, and
\beq               \label{etas}
	\eta_1 = \frac{2 x  - B_1}{\sqrt{4A_1 + B_1^2}}, \quad
	\eta_N = \frac{2/r -B_N}{\sqrt{4A_N + B_N^2}},
\eeq	
  the quantities $A_1$ and $B_1$ are defined in \rf{A1B1}, $A_N$ and $B_N$ in \rf{ANBN}. The first
  term in \rf{Dphi_C} is the relativistic shift of the angle $\phi$ gained on the half-revolution from 
  the aphelion A to the perihelion P, and all the rest corresponds to the motion from P to a given 
  value of $r$. 
  
  Our next task is to single out the Newtonian part of the relativistic term in \rf{Dphi_C}, bearing
  in mind that we need only first-order corrections to $\phi$, expected to be of the order $10^{-8}$ 
  according to \rf{orders}. In particular, since the square root involved in \rf{Dphi_C} is a small 
  quantity, it can be calculated in Newtonian terms, as $\sqrt{A_N + B_N/r - 1/r^2}$.
  
  More generally, to compare relativistic and Newtonian expressions, we should use directly measurable 
  quantities which can be further used in the framework of any theory. These are the radius $r$ 
  (equal to the circumference around the Sun divided by $2\pi$), the azimuthal angle $\phi$ counted  
  from point A in Fig.\,1, and the time $t$ whose Newtonian version is identified with the time measured 
  by a distant ($r \to \infty$) observer and used in the metric \rf{sph}. The initial velocity 
  $v_0 = r_E (d\phi/dt)$ at point A is also such a measurable quantity.   
    
  Then we obtain for the relativistic integrals of motion in terms of the Newtonian ones:
\bearr               \label{DEDL}
         2{\tilde E} := E^2 - 1 = 2E_m + E_m^2 \approx 2 E_N + v_0^4 + \frac{k}{r_E^2},
\nnn
         L^2 \approx L_N^2 \bigg(1 + v_0^2 + \frac{2\mu}{r_E}\bigg),
\ear  
  whose orders of magnitude are $\tilde E \sim 10^{-8}$, $L^2 \sim 10^8$ km. In obtaining this,
  we took into account that at probe motion near point A we have $ds^2 = f(r_E) dt^2 -r_E^2 d\phi^2$.
  Furthermore, for the quantities appearing in \rf{Dphi_C}, we get
\bearr             \label{xi1}
  		A_1 \approx \frac{2 \tilde E}{L_N^2}(1 - \xi_1),\qq
  		B_1 \approx \frac{2 \mu}{L_N^2}(1 - \xi_1),
\nnn
  		\xi_1 := v_0^2 + \frac{2\mu}{r_E} + \frac{2\mu^2(\beta-\gamma-1)}{L_N^2},
\ear  
  where $A_1 \sim 10^{-16}\, \rm km^{-2}$, $B_1 \sim 10^{-8}\, \rm km^{-1}$, $\xi_1 \sim 10^{-8}$. 
  
  Let us evaluate $\Delta\phi$ at point C where $r = r_V$, and $\phi$ is close to its 
  Newtonian value $\phi_N = 3\pi/2$. 
  Accordingly, the last, Newtonian term in \rf{Dphi_C} is equal to $ -\pi/2$. On the other hand,  
  we can recall that at the lower limit, $r = r_P$, according to \rf{eta-pm}, the relativistic quantity 
  $\eta_1 = 1  \then  \arccos\eta_1 =0$. At the upper limit $r = r_V$, we have 
  $\eta_N = 0 \then \arccos\eta_N =\pi/2$, hence $\eta_1$ is a small quantity,
  and $\arccos \eta_1 \approx \pi/2 - \eta_1$. Also, the term in \rf{Dphi_C} with a square
  root is zero at $r = r_P$. Summarizing all that, we can rewrite \rf{Dphi_C} as follows
  with appropriate accuracy:
\bearr                  \label{Dphi}
		\Delta\phi = \frac{3\pi}{2}\,\frac{\mu(2 + 2\gamma -\beta)}{a(1-e^2)}
				- \eta_1(r_V) 
\nnn		\qq		
				+ (\gamma+1)\mu \sqrt{A_N + B_N/r_V -1/r_V^2},
\ear   
  where $A_N$ and $B_N$ are given by \rf{ANBN}, and
\beq                 \label{eta1}
		\eta_1(r_V) = \frac {B_N (r_V \xi_1 - \mu)}{r_V \sqrt{4A_N + B_N^2}}.
\eeq  
  Thus we have an analytical expression for the relativistic azimuthal angle shift at point C.

\subsection{Probe motion near Venus and later}  
\begin{figure}
\centering
\includegraphics[width=4.5cm]{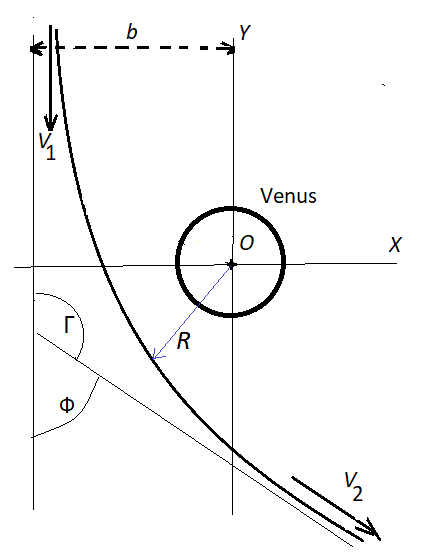} 
\caption{\small 
		The probe motion near Venus}
		\label{fig2}
\end{figure}  
  In the case where the assist takes place at point C with $\phi = 3\pi/2$ in the Newtonian 
  approximation, the calculated values of the probe velocity components at C are
\bearr         \label{v1_xy}
		v_{1x} = v_V = 35.02530368\ {\rm km/s}, 
\nnn
		v_{1y} = -9.68976496\ {\rm km/s}
\ear
  in the Cartesian coordinates of Fig.\,1, in the Sun's RF. It is convenient to 
  consider the probe motion in Venus's gravitational field in its own RF, using the Cartesian 
  coordinates parallel to those of Fig.\,1, since Venus's orbital velocity $v_V$ at point C is $x$-directed. 
  Thus in Venus's RF the probe velocity has the components\footnote
  		{We denote by capital letters $R, V, X, Y, \Phi$ the corresponding quantities in Venus's RF
  		in the coordinate system centered at Venus's center (Fig.\,2).}
\beq           \label{V1_xy}  
		V_{1X} = 0,\qq
		V_{1Y} = -9.68976496\ {\rm km/s}.
\eeq  
  We consider the probe motion in Venus's RF with the initial data \rf{V1_xy} and 
\beq             \label{b}  
		X_1 = -b, \qq  Y_1 = \infty,
\eeq  
  where $b$ is the impact parameter (Fig.\,2), a free parameter of the problem.
  The characteristic length scale at this motion is $\sim 10^3 - 10^4$ km, 
  much smaller than that of the interplanetary motion ($\sim 10^8$ km), which allows us to put 
  $Y_1 = \infty$ at the beginning of the process. Our goal is to find the final velocity 
  components $v_{2x}$ an $v_{2y}$ to be used as initial data for the probe's further motion 
  in the solar gravitational field. This approach, called the single instant hyperbola (SIH) 
  method \cite{yef-20}, corresponds to the approximation in which the sphere of a planet's
  gravitational influence (the Hill sphere) is infinitesimally small from the viewpoint of the 
  probe's interplanetary motion, making it possible to consider the GA as an instantaneous event. 
  
  Under the conditions \rf{V1_xy} and \rf{b}, in full similarity with \eqs \rf{eq-N}--\rf{ANBN}, it is 
  easy to calculate the total angle $\Gamma$ covered by the radius vector $R$ of the probe 
  (see Fig.\,2) at its hyperbolic motion in Venus's gravitational field:
\bear             \label{Gam}
		\Gamma \eql 2 \int_{R_{\min}}^\infty \frac {d Z}{\sqrt{A_V + B_V Z - Z^2}}
\nn		
			      \eql 2 \arccos \frac{-2/R + B_V}{\sqrt{4A_V + B_V^2}}\Bigg|_{R_{\min}}^\infty, 
\ear 
  where $R_{\min}$ is the minimum distance of the probe from the center of Venus, $Z =1/R$, and
\beq  
		A_V = \frac{2E_V}{L_V^2} = \frac 1{b^2}, \qq 
		B_V = \frac{2\mu_V}{L_V^2} =  \frac{2\mu_V}{b^2 V_1^2};
\eeq
  here $V_1 = -V_{1Y}$ is the absolute value of the initial velocity of the probe, 
  $E_V = V_1^2/2$ and $L_V = b V_1$ are the conserved energy and angular momentum per
  unit mass of the probe, respectively, and 
\beq
		\mu_V = GM_V \approx 3.24872\ten{5}\, \rm\frac{km^3}{s^2},
\eeq  
  $M_V \approx 4.8675\ten{24}$\,kg being Venus's mass. The value of $R_{\min}$ is obtained 
  from the condition $\dot R =0$ in the integral of the equations of motion similar to \rf{int-N}:
\beq
		R_{\min} = \frac{1}{V_1^2} \Big(-\mu_V + \sqrt{\mu_V^2 + b^2 V_1^4}\Big).
\eeq  
  As a result, \eqn{Gam} leads to the expression confirming the general relations for hyperbolic 
  motion in a Newtonian gravitational field\cite{LL-meh}   
\beq
		\Gamma = 2 \arccos \frac 1{e_V}, \qq  e_V = \sqrt{1+ \frac {V_1^4 b^2}{\mu_V^2}}.
\eeq  
 where $e_V$ is the eccentricity $e_V$ of the probe's hyperbolic orbit near Venus.
 The total deflection angle $\Phi$ is related to $\Gamma$ as
\beq
		\Phi = \pi - \Gamma = 2 \arcsin \frac 1{e_V}.
\eeq  
  It is now straightforward to find the components of the probe final velocity ${\vec V}_2$ 
  in Venus'e RF since, due to energy conservation, $V_2 = |{\vec V}_2| = V_1$. Thus,
\beq
		V_{2X} = V_1 \sin \Phi, \qq    V_{2Y} = - V_1 \cos \Phi
\eeq   
  (the minus sign is here due to the initial negative direction of the velocity along the $Y$ axis).
  
\def\tall{\raisebox{5pt}{\mathstrut}}
\begin{table*}[t]
\centering
\caption{\small
		Probe motion parameters at and after the gravity assist at different values of the impact
		parameter $b$: the minimum distance from the center of Venus $R_{\min}$,
		the deflection angle $\Phi$ in Venus's RF, the velocity components $v_{2x}$ and $v_{2y}$
		in Sun's RF after the assist, the orbit excentricity $e_2$, the aphelion radius $r_{2+}$ 
		after the assist, the angle $\phi_2(b)$ given by \rf{tan}, the travel time $t_{2+}$
		from the orbit of Venus to the aphelion, and the corresponding sensitivities 
		$dr_{2+}/db$, $\Delta \ell_{\tan}/\Delta b$ and $dt_{2+}/db$ to variations of $b$.}
\bigskip				
\begin{tabular}{|c||c|c|c|c|c|}
\hline
	     $b$ (km)  				&  10000 & 12000 & 14000 & 16000 & 18000  \wide \\
\hline
		$R_{\min}$ (km) 		& 7121.61 & 9028.81 & 10961.2 & 12909.8 & 14869.5 \tall
	\yy	
		$\Phi$ (rad)			& 0.666227 & 0.56145 & 0.484585 & 0.42595 & 0.379819 
	\yy
		$v_{2x}$ (km/s)		& 41.0138 & 40.1843 & 39.5392 & 39.029 & 38.6178 
	\yy
		$v_{2y}$ (km/s)		& -7.6177 & -8.20223 & -8.57416 & -8.82395 &  -8.99919 
	\yy
		$e_2$          		     & 0.450155 & 0.414992 & 0.389411 & 0.370431 & 0.356034 
  	\yy	
    		$r_{2+}$ ($10^6$ km)   & 269.84   &  243.47 & 225.84 &  213.415  &  204.271 
     \yy  	
		$dr_{2+}/db$		     & -16308.1  & -10606.2 & -7303.74  & -5274.42 & -3958.22  
    	\yy	
	     $\phi_2$ (rad)	           & 2.54024  &  2.4374  & 2.35258  &  2.28159  &  2.22147
    	\yy	
	$\Delta \ell_{\tan}/\Delta b$ & -15346.2 & -11319.4 & -8731.82 & -6950.56 & -5664.2
     \yy
          $t_{2+}$ (days)          & 236.271 & 205.297 & 184.508  & 169.698  & 158.65
     \yy
     	    $dt_{2+}/db$ (s/km)    & -1658.40 & -1076.59 & -747.845 & -546.146 & -416.512
      \\
\hline
\end{tabular}
\end{table*}

  In Sun's RF the velocity components are
\beq
		v_{2x} = V_{2X} + v_V, \qq  v_{2y} = V_{2Y}.
\eeq   
  These components must be used as the initial data for calculating the probe motion in Sun's 
  gravitational field after gravity assist at point C, to be considered in the framework of Newtonian 
  gravity. The conserved quantities (integrals of motion) $E_{N2}$ and $L_{N2}$ are
\bear
		E_{N2} \eql \Half \Big(v_{2x}^2 + v_{2y}^2\Big)-\frac {GM_\odot}{r_V}, 
\nn		
		L_{N2} \eql  r^2 \dot\phi = r_V\, v_{2x},
\ear  
  both being functions of the impact parameter $b$. The corresponding geometric parameters
  (the eccentricity $e_2$, the orbital parameter $p_2$, the major semiaxis $a_2$ and the 
  aphelion and perihelion radii $r_{2\pm}$) of the resulting probe orbit are
\bearr
		e_2 = \sqrt{1+ {2 E_{N2} L_{N2}^2}/{\mu^2}}, 
\nnn
		p_2 = \frac{L_{N2}^2}{\mu}, \quad     a_2 = \frac{\mu}{2 E_{N2}}, \quad
		r_{2\pm} = \frac{p_2}{1 \mp e_2},
\ear
  where, as before, $\mu = GM_\odot \approx 1.4777$\,km in units where $c=1$, or, in 
  usual units more appropriate for calculations in Newtonian theory, 
\beq
		\mu \approx 1.327461\ten{11}\,\rm\frac{km^3}{s^2}.  
\eeq  

   Of interest for us is the sensitivity of the probe aphelion position at changes of the impact 
   parameter $b$. In addition to the radius $r_+(b)$, we should determine the aphelion
   location in the tangential direction, which can be characterized by the angle
\bearr          \label{tan}
		\phi_2(b) = \pi - \arccos\frac {p_2-r_V}{e_2 r_V}, 
\ear 
   covered during probe motion from the GA position (crossing Venus's orbit, point C in Fig.\,1) 
   to the aphelion, calculated according to \rf{phi-N}. 
   The tangential shift $\Delta \ell_{\tan}$ of the aphelion due to a small change of $b$ ($\Delta b$)
   is then estimated as  
\beq              \label{l-tan}
		\Delta \ell_{\tan} = r_{2+}(b) \Delta\phi_2 (b), 
\eeq   
  where $\Delta \phi_2 (b)$ is the increment of $\phi_2 (b)$ due to $\Delta b$. Both radial and 
  tangential shifts of the aphelion due to small $\Delta b$ are illustrated in Fig.\,1 as the position 
  difference between the points $\rm A'$ and $\rm A''$. 
 
  One more quantity of interest is the travel time $t_{2+}$ from the assist position to the aphelion, 
  which, according to \eqn{int-N}, is given by
\bearr             \label{t-aph}
		t_{2+} = \int_{r_V}^{r_{2+}} \frac{r\, dr}{\sqrt{2r^2 E_{N2} + 2\mu r - L_{N2}^2 }}
\nnn  \qq
		= \Bigg[ \frac {\sqrt{2r^2 E_{N2} + 2\mu r - L_{N2}^2 }}{2E_{N2}}
\nnn	\ \ 	
		- \frac{\mu}{|2E_{N2}|^{3/2}} 
		\arcsin \frac {\mu + 2 r E_{N2} }{\sqrt{\mu^2+2E_{N2} L_{N2}^2 }}\Bigg]_{r_V}^{r_{2+}} .	
\ear
  
\subsection{Numerical estimates}  

  Let us estimate the relativistic correction \rf{Dphi} according to the data \rf{N_data}. 
  A term by term calculation of this correction in \eqn{Dphi} gives
\bear                      \label{Dphi-N}
		\Delta\phi \al \approx \al 6.432\ten{-8} (2+2\gamma -\beta) + 5.084 \ten{-8} 
\nnn  		
		+ 9.871\ten{-8} (\gamma - \beta) + 3.774 \ten{-9} (\gamma+1)   
\nn
		\al \approx \al (1.83 - 1.63 \beta + 2.31 \gamma) \ten{-7}.  		
\ear		
  At motion near point C, this $\Delta\phi$ shifts the probe trajectory to the right by the
  distance $r_V \Delta\phi$, thus decreasing the impact parameter $b$ (see Fig.\,2) by
\beq              \label{Db}
		\Delta b \approx - (19.83 - 17.64\beta + 25.01 \gamma)\ {\rm km}.
\eeq    
  In particular, in the case of GR with $\beta = \gamma =1$, we obtain $\Delta b \approx 27.20$\ km. 

  This change in $b$ results in a substantial change in the parameters of probe motion at stage 2,
  as follows from Table 1 that presents these parameters for a few values of $b$.
  
  In particular, the table shows that if $b$ changes by 1 km from its value of 10000 km, the aphelion 
  radius changes by more than 16000 km, and the aphelion itself ``moves aside'' by more than 
  15000 km. Then the relativistic correction \rf{Db} shifts the aphelion point by more than 
  500 thousand kilometers. The parameters $\beta$ and $\gamma$ are known to be close to unity 
  (their GR values) up to $\sim 10^{-4}$ \cite{will-14}, hence to improve the knowledge of these 
  quantities using the GA method as described here, it is necessary to determine the aphelion position 
  better than up to $\sim 50$ km (provided that other sources of error are excluded), which 
  looks quite possible.  
  
  The probe travel time is also highly sensitive to the values of $b$: as follows from the last line 
  of the table, the whole relativistic correction to $t_{2+}$ is about 45000 seconds for $b=10000$ km, 
  therefore, to improve the knowledge of $\beta$ and $\gamma$ one should fix the time when 
  the probe reaches its aphelion up to a few seconds.
  
\section{Concluding remarks}  

  We have demonstrated that the gravity assist maneuver can be regarded as an efficient tool
  for measuring the parameters of metric theories of gravity. Using the Eddington parameters 
  $\beta$ and $\gamma$, which characterize vacuum spherically symmetric gravitational fields in such
  theories, we have estimated the sensitivity of probe trajectories to variation in $\beta$ and $\gamma$ 
  on the basis of the flight model considered in \cite{yef-20, yef-21}. We found that variations of the 
  order of $10^{-4}$ in $\beta$ or $\gamma$ (corresponding to the accuracy of the present knowledge 
  of their values) lead to shifts of about 50 km in the probe's aphelion position. Thus the GA maneuver 
  significantly amplifies small variations in the trajectory parameters thus making the method 
  exceedingly sensitive.
  
  As follows from Table 1, the method sensitivity rapidly grows as the impact parameter $b$ gets smaller, 
  but we evidently cannot approach too close to the planet to avoid touching its atmosphere:
  in the case of Venus, $R_{\min}$ should not be smaller than $\approx 6500$ km.   
  
  On the other hand, according to \cite{yef-20}, an even larger sensitivity may be gained by 
  observing the probe arrival point back on the Earth's orbital radius; it seems, however, that
  such arrival is harder to precisely observationally fix than the aphelion point.
  
  For our demonstration purposes it was sufficient to adhere to spherically symmetric metrics and 
  circular planetary orbits, and to use some realistic approximations from the viewpoint of the 
  currently achievable measurement accuracy. In particular, the probe motion at the GA stage 
  and afterwards was studied in the framework of the Kepler problem in Newtonian gravity, while 
  small relativistic corrections appreciably affect only the GA impact parameter $b$. However, in 
  possible future practical experiments it will be evidently necessary to take into account all 
  significant factors affecting the probe motion, such as ellipticity of planetary orbits,
  perturbations from the gravitational field of other planetary and tthe solar wind, and subtle 
  details of the GA itself beyond the single instant hyperbola approximation.

\Acknow{The authors are grateful to Alexander P. Yefremov for fruithful discussions.}

%
%

\end{document}